%% file: Synchronization-with-Maxima-arXiv.tex
\documentclass[12pt,a4paper]{article}
\usepackage{times}                                        %
\usepackage{amsfonts,amsmath,amssymb}                     %
\usepackage[a4paper]{geometry}                            %
\usepackage{fancyhdr}                                     %
\usepackage{color}                                        %
\usepackage[pdftex]{hyperref} 
\usepackage{graphicx}
\hypersetup{%
  a4paper,
  pdfpagemode={UseNone},
  bookmarks=false,
  breaklinks=true,
  colorlinks=true,
  linkcolor=blue,    
  urlcolor=blue,     
  citecolor=blue
}                                            %
{\ \rule{0.5em}{0.5em}}                                   %
\newcounter{ejmtFirstpage}                                %
\usepackage[dvipsnames]{xcolor}
\usepackage{graphicx}
\usepackage{animate}
\usepackage{tikz}
\usetikzlibrary{calc,patterns,angles,quotes,arrows}
\usepackage{pgfplots}
\pgfplotsset{compat=newest}
\begin{document}
%
%
\title{Synchronization of dynamical systems: an approach using
a Computer Algebra System}%
%
%
\author{\begin{tabular}{cc}
\textit{Guillermo D\'avila} & \textit{Antonio Morante and Jos\'e A Vallejo} \\
\small{\texttt{davila@gauss.mat.uson.mx}} & \small{\texttt{amorante,jvallejo@fciencias.uaslp.mx}} \\
Departamento de Matem\'aticas & Facultad de Ciencias \\
Universidad de Sonora & Universidad Aut\'onoma de San Luis Potos\'i\\
Hermosillo (Son) CP 83000 & San Luis Potos\'i (SLP) CP 78290 \\
M\'exico & M\'exico \end{tabular}
}%
%
\date{}                                                   %
\maketitle                                                %
%
%
\begin{abstract}
The synchronization between two dynamical systems is one of
the most appealing phenomena occuring in Nature. Already
observed by Huygens in the case of two pendula, it is a
current area of research in the case of chaotic systems,
with numerous applications in Physics, Biology or Engineering.
We present an elementary but detailed exploration of the theory
behind this phenomenon, including some graphical animations, 
with the  aid of the free CAS Maxima, but the code can be 
easily ported to other CASs. The examples used are the Lorentz
attractor and a pair of coupled pendula because these are 
well-known models of dynamical systems, but the procedures 
are applicable to any system described by a system of 
first-order differential equations.
\end{abstract}%
%
\thispagestyle{fancy}                                     %
%
%

\section{Introduction}

Synchronization is, broadly speaking, a process in which two or more systems interact with 
each other resulting in a joint evolution on some of their dynamical properties. 
These systems are capable of adjusting its pace, `influenced one by the other', 
showing the same behaviour over time.

Synchronization phenomena pervades our daily lives. Many of our bodies physiological functions are 
synchronized to the day-night cycle (circadian rhythm); thousands of pacemaker cells in the 
sinoatrial node, fire at unison in order to maintain the regular beats of our hearts; in many 
countries, thousands of fireflies reunite at night along riverbanks and synchronize their flashes 
in an amazing spectacle that has been noticed and reported for over three centuries; laser beams 
are also examples of perfect synchronization of trillions of atoms. As is apparent from these 
examples, synchonization phenomena occur mainly between self-sustained oscillators, through some 
kind of coupling between them.

In this paper, we explore some scenarios for the synchronization of dynamical systems by using 
the  Computer Algebra System (CAS) Maxima \cite{5b} and its graphical interface wxMaxima, 
emphasizing the 
synchronization of \emph{chaotic systems} since, from a pedagogical viewpoint, this approach
can lead to a better understanding of both topics, chaos and synchronization. We have chosen 
Maxima because it is free, open software released under the General Public License and it is 
available for most used platforms, i.e., Linux$^{\copyright}$, Windows$^{\copyright}$,
Android$^{\copyright}$ and Mac OS$^{\copyright}$. Moreover, the simulations with Maxima 
presented here can also be implemented in other CAS like Mathematica$^{\copyright}$, 
Maple$^{\copyright}$, and Matlab$^{\copyright}$, without major effort (however, in order to use the interactive buttons in the pdf file, Acrobat$^{\copyright}$ Reader is needed).

The paper is organized as follows. In Section \ref{sec2}, we present some generalities about 
dynamical systems and introduce some Maxima commands to deal with them. In Section \ref{sec3}, 
some basic ideas about chaotic dynamical systems are described and we explore chaos through a 
paramount example, namely, the Lorenz system. Synchronization of coupled systems is addressed in 
Section \ref{sec4} and we show how it is possible to synchronize chaotic ones, which is one of the 
main purposes of this paper, with two different techniques: the replacement and the master-slave
synchronizations. Finally, in Section \ref{pendula} we illustrate how to synchronize 
two pendula, along the lines of Huygnes observations on the subject. Some additional theoretical 
results on stability and Lyapunov functions are treated in the appendix.

\section{Dynamical Systems}\label{sec2}

We use the term \emph{dynamical system} as a synonym of `physical system', such as the ones
satisfying Newton's equations: If a certain physical system depends on $k$ parameters (generalized 
coordinates in Mechanics, temperatures and pressures in Thermodynamics, and so on), we can
consider $\mathbb{R}^k$ as the \emph{evolution space}, and think that the behavior of the system
is described by the curve $\mathbf{x}(t)=(x_1(t),\ldots,x_k(t))$ satisfying the equations
\begin{equation}\label{eq3}
m\ddot{\mathbf{x}}(t)=\mathbf{F}(\mathbf{x}(t),\dot{\mathbf{x}}(t))\,,
\end{equation}
where $m$ is the mass, and $\mathbf{F}:\mathbb{R}^k\times\mathbb{R}^k\to\mathbb{R}^k$ is a (sufficiently regular)
mapping characterizing the \emph{force} acting on the system\footnote{We will not consider here
time-dependent forces, as these are quite rare in Physics (contrary to what happens in
Engineering).}.

Now, if the evolution space is $\mathbb{R}^k$ with coordinates $(x_1,\ldots,x_k)$, we will
consider the doubled space $\mathbb{R}^{2k}=\mathbb{R}^k\times\mathbb{R}^k$ and call it the
\emph{phase space}. The coordinates in this space will be the $(x_1,\ldots,x_k)$ \emph{and}
the new ones $(y_1,\ldots,y_k)$ corresponding to the second factor in $\mathbb{R}^{2k}=\mathbb{R}^k\times\mathbb{R}^k$. That is, a point in the phase space will
be denoted 
$$
(\mathbf{x},\mathbf{y})=(x_1,\ldots,x_k,y_1,\ldots,y_k)\in\mathbb{R}^{2k}\,,
$$
and a curve in phase space will be parametrized as $(\mathbf{x}(t),\mathbf{y}(t))$.

To make use of our brand new phase space, we define a new curve containing the values of
the derivatives at each instant:
$$
\mathbf{y}(t)=\dot{\mathbf{x}}(t)\,,
$$
and consider the following system of equations in phase space, completely equivalent to
\eqref{eq3}:
\begin{equation}\label{eq4}
\begin{cases}
\dot{\mathbf{x}}(t)= \mathbf{y}(t)\\
\dot{\mathbf{y}}(t)= \dfrac{1}{m}\mathbf{F}(\mathbf{x}(t),\mathbf{y}(t))\,.
\end{cases}
\end{equation}
If we define $\mathbf{f}:\mathbb{R}^{2k}\to\mathbb{R}^{2k}$ by
$$
\mathbf{f}(\mathbf{x},\mathbf{y})=
\left(\mathbf{y},\dfrac{1}{m}\mathbf{F}(\mathbf{x},\mathbf{y})\right)\,,
$$
we see that any Newtonian system can be described through a vector field on phase space.
This motivates the following definition: \emph{a dynamical system is just a vector field
$\mathbf{f}$ on a suitable phase space $\mathbb{R}^n$}. Denoting collectively the 
coordinates in phase space by $\mathbf{z}$, this is the same as 
the system of first-order differential equations
$$
\dot{\mathbf{z}}(t)=\mathbf{f}(\mathbf{z}(t))\,.
$$

Time for an example. Consider a plane pendulum, that is, a mass $m$ hanging on a thin rod
(of negligible weight) of length $\ell$,
separated a small distance from the local vertical and left alone, evolving under the
action of gravity. The parameter in this case is the angle the bar forms with the local
vertical passing through the hanging point, which we will denote $\theta$, and the total
force is the sum of gravity $\mathbf{g}$ and the tension $\mathbf{T}$ along the rod.

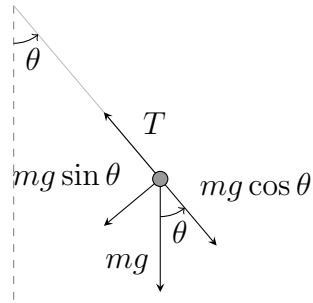
\begin{figure}[h]\centering
\begin{tikzpicture}
    \pgfmathsetmacro{\Gvec}{1.5}
    \pgfmathsetmacro{\myAngle}{40}
    \pgfmathsetmacro{\Gcos}{\Gvec*cos(\myAngle)}
    \pgfmathsetmacro{\Gsin}{\Gvec*sin(\myAngle)}

    \coordinate (centro) at (0,0);
    \draw[dashed,gray,-] (centro) -- ++ (0,-4) node (mary) [black,below]{$ $};
    \draw[gray!50] (centro) -- ++(270+\myAngle:3) coordinate (bob);
    \pic [draw, ->, "$\theta$", angle eccentricity=1.5] {angle = mary--centro--bob};
    \draw [-stealth] (bob) --node[midway,above right]{$T$} ($(bob)!\Gcos cm!(centro)$);
    \draw [-stealth] (bob) -- ($(bob)!-\Gcos cm!(centro)$)
      coordinate (gcos)
      node[midway,above right] {$mg\cos\theta$};
    \draw [-stealth] (bob) -- ($(bob)!\Gsin cm!90:(centro)$)
      coordinate (gsin)
      node[midway,above left] {$mg\sin\theta$};
    \draw [-stealth] (bob) -- ++(0,-\Gvec)
      coordinate (g)
      node[near end,left] {$mg$};
    \pic [draw, ->, "$\theta$", angle eccentricity=1.5] {angle = g--bob--gcos};
    \filldraw [fill=black!40,draw=black] (bob) circle[radius=0.1];
\end{tikzpicture}\caption{A vertical pendulum}\label{fig2}
\end{figure}
Considering the diagram in Figure \ref{fig2} we see that the radial forces (the tension $T$ and the component $mg\cos\theta$ of gravity)
cancel each other, so the net force is the tangential component $mg\sin\theta$. Newton's
equations read (after canceling the mass term)
\begin{equation}\label{eq5}
\ell\ddot{\theta} =-g\sin\theta\,,
\end{equation}
with the minus sign taking into account that the force is directed toward the vertical passing through 
the hanging point.
Defining our phase space as $\mathbb{R}^2$ and mimicking what we did to get \eqref{eq4}, we
introduce the vector field
\begin{equation}\label{eq7}
\mathbf{f}(x,y)=(y,-\frac{g}{\ell}\sin x)\,.
\end{equation}
Now we see that the Newtonian description of the pendulum, that is, finding the expression of the 
angle as a function of time $\theta (t)$, is the same as finding the integral curves of the dynamical
system on the phase space $\mathbb{R}^2$ given by
$\dot{\mathbf{x}}(t)=\mathbf{f}(\mathbf{x}(t))$.

Regarding this example, let us remark that \eqref{eq5} does not admit a closed solution in
terms of elementary functions (the explicit solution can be written in terms of the complete
and incomplete elliptic integrals of the first kind). However, in the regime of small-amplitude
oscillations $\sin\theta \simeq \theta$, and equation \eqref{eq5} reduces (aside from a scale
factor) to the linear equation $\ddot{x}(t) + x(t)=0$, whose solutions are superpositions of
sine and cosine waves.

Of course, a numerical integration is always possible. The Maxima package \texttt{drawdf}
implements the fourth-order Runge-Kutta method for integrating systems of first-order
ordinary differential equations, so it can provide accurate plots of the integral curves.
For instance, taking $g=10\,m/s^2$ and $\ell =2\,m$ we could issue the commands
\noindent
\begin{verbatim}
(%i1) load(drawdf)$
(%i2) wxplot_size=[600,400]$
(%i3) wxdrawdf([y,-5*sin(x)],
solns_at([0,-5],[0,-3],[0,5],[%pi-0.0001,0]));
(%o3)
\end{verbatim}
$$
\includegraphics[scale=0.5]{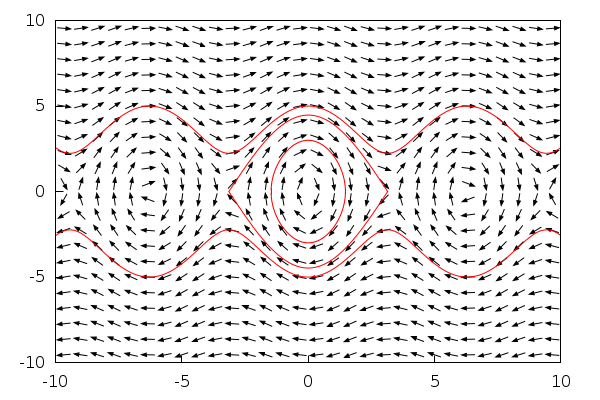} 
$$

Notice that the point $(x,y)=(\pi ,0)$ is a \emph{fixed point} of the system \eqref{eq7},
that is, the constant curve staying on it $\mathbf{x}(t)=(\pi,0)$ is a valid solution. Thus,
if we were to write \texttt{solns\_at([\%pi,0])} in the previous command, we would see just a point
(the trajectory passing through that point is the point itself), this is the reason for taking
$\pi -0{.}0001$ to illustrate the oval-shaped trajectories in phase space. Another fixed point
corresponds to $(0,0)$. The corresponding constant curves are called \emph{equilibrium solutions},
and the difference between the two equilibrium solutions $(\pi,0)$ and $(0,0)$ has a physical
interpretation in terms of \emph{stability}. The equilibrium $(\pi,0)$ corresponds to the mass
staying upwards, in a vertical position, and it is \emph{unstable} in the sense that any small perturbation will lead the mass away from this position. This can be seen in the phase diagram
in \texttt{(\%o3)} above by noticing that the small arrows representing the vector field will
take any point close to $(\pi,0)$ upwards to the right (if the point were to the right of
$(\pi,0)$) or downwards to the left (if the point were to the left of $(\pi,0)$). On the contrary,
$(0,0)$, which corresponds to the mass hanging on at the lowest position along the vertical
through the hanging point, is a \emph{stable} equilibrium: The arrows representing the field
clearly shows that any departure from this equilibrium will result in going around it tracing
small circles in phase space, which correspond to small amplitude oscillations around the
vertical, staying close to the equilibrium (values of $\theta$ close to zero).

As an exercise, the reader can try to interpret physically the wavy lines in the phase 
diagram \texttt{(\%o3)}\footnote{Here is a hint: The variable $\theta$ is an angle, so we should
work in a phase space defined modulo $2\pi$, which geometrically results in a cylinder. Try to visualize the wavy line on this cylinder: Does it close on itself?}.

\section{Chaotic Systems}\label{sec3}
Now we turn our attention to a particular class of dynamical systems, those whose behavior
(under certain conditions) can be called \emph{chaotic}. What does this mean? Following
Strogatz \cite[Section~9.3]{9} (see also \cite{1}), we say that:
\begin{quote}
\emph{Chaos} is aperiodic long-term behavior in a deterministic system that exhibits sensitive 
dependence on initial conditions.
\end{quote}

Some comments
on this definition are in order: First of all, the aperiodicity condition implies that there exist
trajectories that do not settle down to fixed points or periodic orbits. It is not necessary that \emph{all} the trajectories satisfy this, we only demand that there exist \emph{some} of them
with that property. Moreover, aperiodicity excludes trajectories tending to infinity for very large
values of their parameter. This is so because `staying at $\infty$' can be considered as a kind of periodicity. In this way, we can distinguish between chaotic behavior and instability: the system
described by $\dot{x}(t)=x(t)$ is unstable (the trajectories $x(t)=e^{t}x_0$ blow up when
$t\to\infty$), but it is not chaotic.

Secondly, the system under study must be \emph{deterministic}, that is, devoid of any stochastic
or random parameters. In this way, we can assign the chaotic features to the system itself, and
not to some external, uncontrolled perturbation.

Finally, the sensitive dependence on initial conditions means that nearby trajectories separate exponentially fast. Even two of them starting infinitely close to each other will separate out
after a brief lapse of time, following different, apparently unrelated, paths.

\subsection{Chaos in 1D}
\begin{figure}[h]\centering
\begin{tikzpicture}[scale=2]
\draw[color=gray!50,thin,->] (-1.75,0.87)--(1.75,0.87)node[right,black]{$t$};
\draw[color=gray!50,thin,->] (0,-0.75)--(0,2.75)node[above,black]{$x$};
\draw[thick] (-1.75,1.75)--(1.75,1.75)node[above,black]{$x_2$};
\draw[thick] (-1.75,0)--(1.75,0)node[above,black]{$x_1$};
\draw[domain=-1.75:1.75,smooth,variable=\x,red,thick]  plot ({\x},{1.75/(1+exp(-4*\x))});
\draw[domain=-1.75:1.75,smooth,variable=\x,red,thick]  plot ({\x},{1.75 + 0.15*exp(-\x)});
\draw[domain=-1.75:1.75,smooth,variable=\x,red,thick]  plot ({\x},{-0.15*exp(-\x)});
\end{tikzpicture}
\caption{Trajectories of $\dot{x}(t)=f(x(t))$}\label{fig3}
\end{figure}
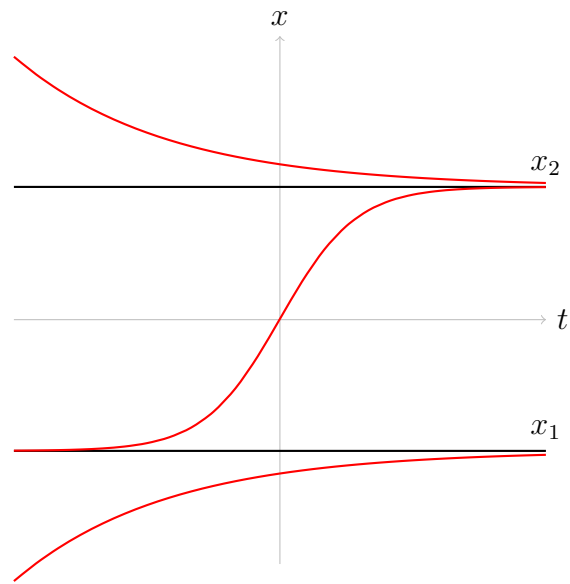

When studying a certain dynamical system, in order to determine its aperiodic long-term
behavior it is essential to know its fixed points (or equilibrium solutions). As we know,
these are the constant solutions $\mathbf{x}(t)=\mathbf{x}^*$, where $\mathbf{x}^*$ is a
solution of the equations
$$
\mathbf{f}(\mathbf{x}^*)=\mathbf{0}\,.
$$
In one dimension (that is, in the case of a dynamical system of the form $\dot{x}(t)=f(x(t))$,
with $x:I\subset\mathbb{R}\to\mathbb{R}$ defined on an open interval $I$ and $f$ at least continuous), 
the geometry is so restrictive that no chaos can appear: Suppose that $x_1$ and $x_2$ are fixed points 
of the system (with $x_1<x_2$, for example), so $f(x_1)=0=f(x_2)$. By continuity, $f$ keeps the same 
sign in the whole interval $\,]x_1,x_2[\,$, say $f(x)>0$. In this case, any solution satisfies
$\dot{x}(t)=f(x(t))>0$ for all the values of $t$ such that $x(t)\in\,]x_1,x_2[\,$, and it is monotonically 
increasing there, being bounded from above by the constant solution $x(t)=x_2$. 
The theorem on existence 
and uniqueness of solutions to differential equations tells us that different trajectories can not cross, so a solution $x(t)$ that at some instant $t_0$ lies in $\,]x_1,x_2[\,$, must approach
asymptotically the equilibrium $x(t)=x_2$. The same reasoning applies to the case $f(x)<0$, this
time giving monotonically decreasing solutions approaching $x(t)=x_1$ asymptotically. Figure
\ref{fig3} illustrates a typical situation: There is no chaos\footnote{This only happens for
\emph{continuous} 1D systems. In the discrete case, there can be chaos even for such a simple
systems as the logistic map, see \cite{6}, for instance.}, every trajectory asymptotically
approaches an equilibrium.
\subsection{Chaos in 2D}
No chaos can happen for dynamical systems in the plane, due to the Poincar\'e-Bendixson
theorem \cite[Section~II.1]{3}. This result states that a plane system 
$\dot{\mathbf{x}}(t)=\mathbf{f}(\mathbf{x}(t))$ given by a \emph{smooth} vector field
$\mathbf{f}:\mathbb{R}^2\to\mathbb{R}^2$, after evolving for a large enough time will reach either:
\begin{enumerate}
\item A fixed point.
\item A periodic orbit.
\item A connected set composed of a finite number of fixed points and trajectories connecting them
(these trajectories connecting fixed points among themselves are called \emph{homoclinic} if they
start and finish at the same fixed point, and \emph{heteroclinic} otherwise).
\end{enumerate}
The example of the pendulum (analyzed in the preceding section) illustrates these possibilities.
There are fixed points (equilibrium solutions) at each $(k\pi,0)$, with $k\in\mathbb{Z}$. Then we
have closed trajectories (periodic orbits) surrounding each equilibrium of the form
$(2k\pi,0)$, while there appear heteroclinic trajectories between those of the form
$((2k-1)\pi,0)$ and $((2k+1)\pi,0)$, as shown with the Maxima commands below.
\begin{verbatim}
(%i4) wxdrawdf([y,-5*sin(x)],soln_arrows=true,
    solns_at([0,-3]),
    point_size=0.5,points_at([0,0],[2*%pi,0],[-2*%pi,0]),
    saddles_at([-%pi,0],[%pi,0])),
    wxplot_size=[600,400];
(%o4)
\end{verbatim}
$$
\includegraphics[scale=0.5]{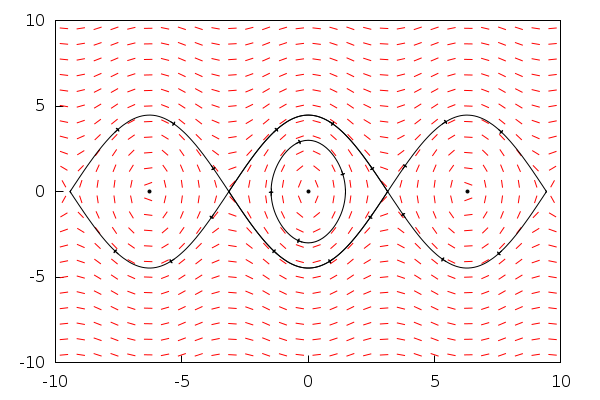} 
$$

The preceding command introduces to the reader three new options of \texttt{drawdf}: One of them is
\texttt{soln\_arrows=true}, which adds some little arrows to the trajectories and omits them
from the vector field (also inverting the colors, for greater clarity). The option \texttt{points\_at}, obviously, draws points with the radius specified in \texttt{point\_size}
at the given places. The last one is
\texttt{saddles\_at}, which instructs Maxima to linearize the equations on a neighborhood of
the specified points, drawing the heteroclinic trajectories arising from them.

Again, the reader is invited to think about the physical meaning of the wavy lines seen in
\texttt{(\%o3)}. Do their existence contradicts the conclusions of the Poincar\'e-Bendixson
theorem? Another exercise (actually related to the previous one): The Poincar\'e-Bendixson theorem
concludes the existence of a \emph{finite} set of equilibria connected by heteroclinic
trajectories, but we have seen that there are an infinity of them (located at
$(2k+1\pi,0)$ for $k\in\mathbb{Z}$), although we have represented only five. What is wrong here?

\subsection{Chaos in 3D}
Chaos is possible for nonlinear systems in 3D. A basic example was discovered by E. N. Lorenz
in 1963 \cite{5}, while working in a simplified mathematical model of atmospheric convection (the vertical circular motion of air masses near the Earth surface).

Lorenz proposed the following system of nonlinear equations, where $b,r,\sigma$ are some
parameters to be determined by experimental data:
\begin{equation}\label{eq8}
\begin{cases}
\dot{x}=-\sigma x +\sigma y \\
\dot{y}=r x -y -x z\\
\dot{z}= xy-b z\,.
\end{cases}
\end{equation}
Notice that this is a clearly deterministic system (although it contains parameters, they are
not random at all). To check that the remaining conditions for chaotic motion are satisfied, 
we compute the trajectory starting at $(x_0,y_0,z_0)=(-8,8,27)$, using $5\,000$ iterations
and the values originally suggested in \cite{5}, $\sigma =10,r=28,b=8/3$:

\begin{verbatim}
(%i5)	s:10.0$
(%i6)	r:28.0$
(%i7)	b:float(8/3)$
(%i8)	data1:rk([s*y-s*x,-x*z+r*x-y,x*y-b*z],
	             [x,y,z],[-8,8,27],[t,0,50,0.01])$
\end{verbatim}

The output of \texttt{rk} is a list \texttt{data1} containing $5\,000$ elements, each one of them 
being a sublist with the format $[t_i,x_i,y_i,z_i]$. To represent the points in 3D space, we must 
delete the first element of each sublist, and this can be achieved by mapping the function
\texttt{rest} (which, by default, deletes the first element of a list and returns the rest, hence the
name) onto \texttt{data1}\footnote{To this end we use a \emph{lambda function}. These are pure 
functions or `functions without a name', defined by giving their action on the argument. For instance, 
the function $x\mapsto x^2$ could be described as the one that takes its argument and squares it, or 
\texttt{lambda([x],x\^{}2)}.}. We call \texttt{pnts} the resulting list of 3D points and plot all of 
them together (using some aesthetic options for the \texttt{draw3d} command that should be self-explanatory). 
\begin{verbatim}	             
(%i9)	pnts1:map(lambda([x],rest(x)),data1)$
(%i10)	set_draw_defaults(
        xrange=[-25,25],yrange=[-25,25],zrange=[0,50])$
(%i11)	wxdraw3d(point_type=none,points_joined=true,color=blue,
	    xlabel="x(t)",ylabel="y(t)",zlabel="z(t)",
	    xtics=10,ytics=10,ztics=10,
	    points(pnts1)),
	    wxplot_size=[600,400];	    
(%o11) 
\end{verbatim}
\begin{center}
\input{Figures/animation1-39.tex} 
\end{center}

We can see that this particular trajectory in its long-term evolution traces out a complicated
figure, called the \emph{Lorenz attractor}, that does not approach a fixed point or a closed
periodic orbit, thus exhibiting aperiodicity. In fact, the trajectory visits the two `wings'
of the Lorenz attractor without following any regular pattern, but always staying inside a bounded
region (see \cite{6} for a proof of this boundedness). Perhaps this is best appreciated with an
animation. To create it, we use the command
\texttt{wxanimate\_draw3d}, whose syntax is similar to that of \texttt{draw3d}, but admits
a new argument: A parameter (denoted \texttt{d} below) taking
values from a user-specified list. The \texttt{wxanimate\_draw3d} will generate a set of
graphics, one for each value of the parameter, and after a lapse for preparing the graphics, will
display all of them sequentially, creating an animation effect. In the command below, we construct
a list of values for the parameter \texttt{d} by going from $1$ to $5\,000$ in steps of $50$,
that is, we will have an animation consisting of $100$ slides (the animation will start simply by clicking on the output screen).

\begin{verbatim}	             
(%i12)	wxanimate_draw3d(d,makelist(i,i,1,length(pnts1),50),
    point_type=filled_circle,point_size=0.2,color=blue,
    xlabel="x(t)",ylabel="y(t)",zlabel="z(t)",
    points(rest(pnts1,d-length(pnts1)))),
    wxplot_size=[600,400];
(%o12)
\end{verbatim}
\begin{center}
\begin{animateinline}[controls=true,poster=last]{6}
\multiframe{40}{i=0+1}{
\input{Figures/animation1-\i.tex}
}
\end{animateinline}
\end{center}
\medskip

The output shown here and in the figures that follow can be played back without Maxima, 
using Adobe Acrobat Reader in any of the platforms where it is available:
MS Windows\texttrademark{}, Linux\texttrademark{} or MacOS\texttrademark{} (alternatives
reported to work with embedded \texttt{pdf} animations in MS Windows\texttrademark{} are
Foxit Reader and PDF-XChange).

It remains to check the sensitive dependence on initial conditions. Again, we do it through
an animation displaying the path of two points with very close initial conditions, $(x_0,y_0,z_0)=(-8,8,27)$ and $(x_1,y_1,z_1)=(-8{.}001,8{.}001,27{.}001)$. The command that
achieves this is a slight variation of the preceding one for two sets of points; the first set
is provided by \texttt{pnts1}, the second set is generated as follows:
\begin{verbatim}
(%i13)	data2:rk([s*y-s*x,-x*z+r*x-y,x*y-b*z],
	         [x,y,z],[-8.001,8.001,27.001],[t,0,50,0.01])$
(%i14)	pnts2:map(lambda([x],rest(x)),data2)$
\end{verbatim}

Now we generate the animation. The two trajectories start close each other, 
but they separate far away after a few iterations (they can even be seen swirling around 
different `wings' of the attractor). We have reduced the number of iterations (to $30$) to get a smaller file.
\begin{verbatim}
(%i15)	wxanimate_draw3d(d,makelist(i,i,1,1500,50),
    point_type=filled_circle,point_size=0.275,color=blue,
    points(rest(pnts1,d-length(pnts1))),
    point_type=filled_circle,point_size=0.275,color=orange_red,
    points(rest(pnts2,d-length(pnts2)))),wxplot_size=[600,400];
(%o15)
\end{verbatim}
\begin{center}
\begin{animateinline}[controls=true,poster=last]{6}
\multiframe{30}{i=0+1}{
\input{Figures/animation2-\i.tex}
}
\end{animateinline}
\end{center}


\section{Synchronization}\label{sec4}
In view of the property of sensitive dependence on the initial conditions that the
Lorenz system presents, it is hard to imagine that two such systems could ever be
synchronized starting from different initial states. Thus, it came as a surprise
when L. M. Pecora and T. L. Carroll showed, in 1990 \cite{8}, how to couple two chaotic 
systems in such a way that their motion, starting from very different initial 
conditions, quickly get synchronized. We will explain their original technique,
called the \emph{replacement synchronization}, as well as another popular one, the
\emph{master-slave synchronization}.

Chaotic synchronization is a very active area of research. Its applications range from securing communications over digital channels to the control of bursting in insulin production by beta
cells. Part of the appeal of the theory is due to the fact that it can be understood (and applied)
with a relative lack of sophisticate mathematics, rendering this an ideal topic for undergraduate
courses.

\subsection{Replacement synchronization}

The technique proposed in \cite{8} considers two copies of the system to be 
synchronized\footnote{The actual technique of Pecora and Carroll is a bit more general,
but here we consider this simplified version for the sake of clarity.},
with different initial conditions on them, and then substitute some of the variables in
one copy by variables pertaining to the other (hence the name). If this substitution is
done only in one of the copies, the resulting procedure is called 
\emph{one-way synchronization}, and it is the only one we will consider here. The 
unchanged system is called the  \emph{driving system}, and the other is the 
\emph{response system}.

There is no general recipe for choosing which variables to replace, but at least one
must appear in a non-linear term. Thus, in the case of the Lorenz system \eqref{eq8}
we could take, for instance,
\begin{equation}\label{eq8b}
\begin{cases}
\dot{x}=\sigma (y-x) \\
\dot{y}=r x -y -x z\\
\dot{z}= xy-b z\\
\dot{x}_r=\sigma (y_r-x_r) \\
\dot{y}_r=r x -y_r -x z_r\\
\dot{z}_r= xy_r-b z_r\,,
\end{cases}
\end{equation}
where the indexed coordinates are those of the response subsystem. Notice that
we have just replaced $x_r$ by $x$ in the two equations containing non-linear terms.

To see why this technique works, we analyze the behavior of the difference between
drive and response states $\boldsymbol\xi (t)=\mathbf{x}(t)-\mathbf{x}_r(t)$, for
large values of the time parameter $t$ \cite{9}. We would like to arrive at the conclusion that
$\lim_{t\to\infty}\Vert\boldsymbol\xi (t)\Vert =0$, as in this case we can say that
the synchronization not only will occur, but moreover it will be \emph{asymptotically stable}.
To this end, we will look for an appropriate \emph{Lyapunov function} (see the Appendix
\ref{Appendix} for some terminology regarding stability, as well as some basic facts about 
Lyapunov functions). Substracting the last three equations in \eqref{eq8b} from the first
three, we get the evolution of $\boldsymbol\xi (t)$:
\begin{equation}\label{eq8a}
\begin{cases}
\dot{\boldsymbol\xi}_1=\sigma (\boldsymbol\xi_2-\boldsymbol\xi_1) \\
\dot{\boldsymbol\xi}_2=-\boldsymbol\xi_2- x\boldsymbol\xi_3 \\
\dot{\boldsymbol\xi}_3=x\boldsymbol\xi_2- b\boldsymbol\xi_3 \,.
\end{cases}
\end{equation}
Multiplying the second equation by $\boldsymbol\xi_2$, the third by $\boldsymbol\xi_3$
and adding them, we arrive at
$$
\boldsymbol\xi_2\dot{\boldsymbol\xi}_2 +\boldsymbol\xi_3\dot{\boldsymbol\xi}_3=
-\boldsymbol\xi^2_2- x\boldsymbol\xi_2\boldsymbol\xi_3
+x\boldsymbol\xi_2\boldsymbol\xi_3- b\boldsymbol\xi^2_3=
-(\boldsymbol\xi^2_2 +b\boldsymbol\xi^2_3)\,.
$$
which is clearly a smooth, negative-definite function.
Now, notice that the left-hand side is a total derivative:
\begin{equation}\label{eq8c}
\boldsymbol\xi_2\dot{\boldsymbol\xi}_2 +\boldsymbol\xi_3\dot{\boldsymbol\xi}_3=
\frac{1}{2}\frac{\mathrm{d}}{\mathrm{d}t}\left( \boldsymbol\xi^2_2
+\boldsymbol\xi^2_3\right)\,,
\end{equation}
so, if we define
$$
V(x,y,z)=\frac{1}{2}\left( \frac{x^2}{\sigma} +y^2+z^2 \right)
$$
we actually have a Lyapunov function: $V(x,y,z)$ is clearly
smooth and positive-definite (recall that $\sigma >0$), it has a unique minimum at $(0,0,0)$, and \eqref{eq8c} along with the
first equation of \eqref{eq8a} shows that, along the trajectories of \eqref{eq8a},
\begin{align*}
\frac{\mathrm{d}}{\mathrm{d}t}V(\boldsymbol\xi_1(t),\boldsymbol\xi_2(t),\boldsymbol\xi_3(t))=&
\frac{1}{\sigma}\boldsymbol\xi_1(t)\dot{\boldsymbol\xi}_1(t)-(\boldsymbol\xi^2_2(t) +b\boldsymbol\xi^2_3(t) )\\
=& -(\boldsymbol\xi^2_1(t) -\boldsymbol\xi_1(t)\boldsymbol\xi_2(t))
-(\boldsymbol\xi^2_2(t) +b\boldsymbol\xi^2_3(t) )\\
=&-(\boldsymbol\xi_1(t) -\frac{1}{2}\boldsymbol\xi_2(t))^2
-(\frac{3}{4}\boldsymbol\xi^2_2(t) +b\boldsymbol\xi^2_3(t) )\,,
\end{align*}
(where, in the last step, we have completed the square) that is,
$$
\frac{\mathrm{d}}{\mathrm{d}t}V(\boldsymbol\xi_1(t),\boldsymbol\xi_2(t),\boldsymbol\xi_3(t))
\leq 0\,.
$$
The Lyapunov theorem tells us that, in this case, $\boldsymbol\xi (t)$ will tend towards the
equilibrium at $(0,0,0)$, so $\lim_{t\to\infty}\Vert\boldsymbol\xi (t)\Vert =0$.

The following Maxima commands implement the replacement technique for synchronizing two
Lorenz attractors, starting from initial conditions at $(5,5,5)$ and $(-5,-5,-5)$.
\begin{verbatim}
(%i16)	data3:rk([s*(y-x),r*x-y-x*z,x*y-b*z,
			s*(yr-xr),r*x-yr-x*zr,x*yr-b*zr],
			[x,y,z,xr,yr,zr],[5,5,5,-5,-5,-5],[t,0,10,0.01])$
(%i17)	data31:makelist([p[2],p[3],p[4]],p,data3)$
(%i18)	data32:makelist([p[5],p[6],p[7]],p,data3)$
(%i19)	wxanimate_draw3d(d,makelist(i,i,1,length(data3),30),
    dimensions=[600,500],view=[60,60],
    xrange=[-20,30],yrange=[-25,35],zrange=[0,50],
    point_type=filled_circle,point_size=0.325,color=orange_red,
    points(rest(data31,d-length(data3))),
    point_type=filled_circle,point_size=0.25,color=blue,
    points(rest(data32,d-length(data3))))
(%o19)
\end{verbatim}
\begin{center}
\begin{animateinline}[controls=true,poster=last]{6}
\multiframe{32}{i=1+1}{
\input{Figures/animation3-\i.tex}
}
\end{animateinline}
\end{center}

\subsection{Master-slave synchronization}

The idea of this method is also very simple: Again, instead of thinking of a single
system evolving from two different initial conditions, consider two identical copies
of the system (along with their respective initial conditions), and relate
them through some coupling function. For the sake of definiteness, suppose that we
work with systems defined on $\mathbb{R}^3$, so they have three degrees of freedom.
Let $\mathbf{x}\equiv (x,y,z)$ the generalized coordinates corresponding to
the first system (the \emph{master}), and $\mathbf{x}_s\equiv (x_s,y_s,z_s)$ those of the second
(the \emph{slave}), both determined in their evolution by the same vector field 
$\mathbf{f}:\mathbb{R}^3\to\mathbb{R}^3$; then, we have
$$
\dot{\mathbf{x}}(t)=\mathbf{f}(\mathbf{x}(t)) \mbox{ and } 
\dot{\mathbf{x}}_s(t)=\mathbf{f}(\mathbf{x}_s(t))\,.
$$ 
There are many possibilities for the coupling function, but a popular
choice consists in taking one linear in the difference $\mathbf{x}-\mathbf{x}_s$,
with coefficients possibly dependent on the master variables \cite{10}, that
is,
$$
\mathbf{k}(\mathbf{x})\cdot(\mathbf{x}-\mathbf{x}_s)=
\begin{pmatrix}
k_{11}(\mathbf{x}) & k_{12}(\mathbf{x}) & k_{13}(\mathbf{x}) \\
k_{21}(\mathbf{x}) & k_{22}(\mathbf{x}) & k_{23}(\mathbf{x}) \\
k_{31}(\mathbf{x}) & k_{32}(\mathbf{x}) & k_{33}(\mathbf{x})
\end{pmatrix}
\cdot
\begin{pmatrix}
x-x_s \\
y-y_s \\
z-z_s
\end{pmatrix}\,.
$$
This coupling function is added to the slave subsystem so as to make it respond to the action
of the master subsystem. Thus, the whole system can be described as
\begin{equation}\label{eq9}
\begin{cases}
\dot{\mathbf{x}}(t)=\mathbf{f}(\mathbf{x}(t)) \\
\dot{\mathbf{x}}_s(t)=\mathbf{f}(\mathbf{x}_s(t)) 
+\mathbf{k}(\mathbf{x})\cdot (\mathbf{x}-\mathbf{x}_s)\,.
\end{cases}
\end{equation}

Let us remark that other synchronization schemes
are possible. For instance, a two-way coupling would propose something like
\begin{equation}\label{eq9b}
\begin{cases}
\dot{\mathbf{x}}_1(t)=\mathbf{f}(\mathbf{x}_1(t)) + \mathbf{k}_1(\mathbf{x}_1-\mathbf{x}_2)\\
\dot{\mathbf{x}}_2(t)=\mathbf{f}(\mathbf{x}_2(t)) +\mathbf{k}_2(\mathbf{x}_1-\mathbf{x}_2)\,,
\end{cases}
\end{equation}
and, of course, there is the option of using non-linear functions instead of $\mathbf{k}_1,\mathbf{k}_2$, etc. We will not deepen into these issues here, restricting ourselves to an application of the two-way coupling to pendula synchronization in Section \ref{pendula}.

The gist of the master-slave synchronization technique is that the difference between states,
$\boldsymbol\xi (t) =\mathbf{x}(t)-\mathbf{x}_s(t)$, will go to zero asymptotically, implying that
the states $\mathbf{x}(t)$ and $\mathbf{x}_s(t)$ will approach each other.
If the regularity
class of $\mathbf{f}$ is at least $\mathcal{C}^1$,
this is an easy consequence of the Mean Value Theorem in several variables and a judicious choice
of the matrix $\mathbf{k}$ \cite{2}: We take it equal to
\begin{equation}\label{choicek}
\mathbf{k}=\mathrm{d}_{\mathbf{x}(t)}\mathbf{f} - \mathbf{H}\,,
\end{equation}
where $\mathrm{d}_{\mathbf{x}(t)}\mathbf{f}$ is the differential of the vector field
$\mathbf{f}$ evaluated along the trajectory $\mathbf{x}(t)$ of the master system, and
$\mathbf{H}$ is a \emph{constant} matrix all of whose eigenvalues have strictly negative real parts
(in the engineering literature, this is known as a \emph{Hurwitz matrix}).
For any fixed $t$, taking the difference between the first and second equations in \eqref{eq9} 
we get\footnote{Notice that we have approximated the Jacobian at an intermediate point 
$\mathbf{z}$ (provided by the Mean Value Theorem) by the Jacobian evaluated at $\mathbf{x}(t)$. 
The assumed  $\mathcal{C}^1$ regularity of $\mathbf{f}$ guarantees that the solutions of the 
approximate system will stay close to those of the original one.} 
\begin{align*}
\dot{\boldsymbol\xi}(t) &=  \frac{\mathrm{d}}{\mathrm{d}t}(\mathbf{x}(t)-\mathbf{x}_s(t)) \\
&= \mathbf{f}(\mathbf{x})-\mathbf{f}(\mathbf{x}_s)-\mathbf{k}\cdot\boldsymbol\xi \\
& \simeq (\mathrm{d}_{\mathbf{x}(t)}\mathbf{f} -\mathbf{k})\cdot \boldsymbol\xi \\
&= \mathbf{H}\cdot\boldsymbol\xi\,.
\end{align*}
The solution of this linear equation for $\boldsymbol\xi$ is
$$
\boldsymbol\xi (t)=e^{t\mathbf{H}}\boldsymbol\xi(0)\,,
$$
and the fact that $\mathbf{H}$ is Hurwitz implies that the exponential is a decaying one,
leading to the asymptotic $\lim_{t\to\infty}\Vert\boldsymbol\xi(t)\Vert =0$ (which not 
only shows that there will be synchronization, but that it will be \emph{asymptotically stable},
that is, the systems will remain synchronized for large times).

Let us apply this technique to the Lorenz system \eqref{eq8}. The differential is
\begin{equation}\label{eq10}
\mathrm{d}_{\mathbf{x}(t)}\mathbf{f} =
\frac{\partial (\dot{x},\dot{y},\dot{z})}{\partial (x,y,z)}=
\begin{pmatrix}
-\sigma & \phantom{-}\sigma & \phantom{-}0 \\
r-z(t) & -1 & -x(t) \\
\phantom{-}y(t) & \phantom{-}x(t) & -b
\end{pmatrix}\,.
\end{equation}

Aside from being Hurwitz the matrix $\mathbf{H}$ is completely arbitrary, but a wise
choice is one that cancels as many non constant terms as possible. This principle would lead to
$$
\begin{pmatrix}
-\sigma & \phantom{-}\sigma & \phantom{-}0 \\
\phantom{-}r & -1 & \phantom{-}0 \\
\phantom{-}0 & \phantom{-}0 & -b
\end{pmatrix}\,.
$$
Of course, nothing guarantees that this matrix satisfies the Hurwitz criterion; indeed, from
\begin{verbatim}
(%i1) eigenvalues(
      matrix([-%sigma,%sigma,0],[r,-1,0],[0,0,-b])
      );
(%o1)	
\end{verbatim}
\begin{align*}
&\left[\left[ -\frac{\sqrt{4\sigma r+\sigma^2-2\sigma+1}+\sigma+1}{2},\frac{\sqrt{4\sigma r+\sigma^2-2\sigma+1}-\sigma-1}{2},-b\right]\right. ,\\
& [1,1,1]\Big]
\end{align*}
we see that the sign of the real part of its eigenvalues depends on the numerical relation
between $r$ and $\sigma$. Thus, we introduce a parameter $p$ in $\mathbf{H}$ to compensate
for this fact. As the $\sigma$ and $r$ variables only appear in the first $2\times 2$
block of $\mathbf{H}$, we place $p$ somewhere in there. For instance, we could choose
$$
\mathbf{H}=\begin{pmatrix}
-\sigma & \phantom{-}\sigma & \phantom{-}0 \\
r+p & -1 & \phantom{-}0 \\
0 & \phantom{-}0 & -b
\end{pmatrix}\,,
$$
and we would be ready to study for which ranges of values of $p$ we get a Hurwitz matrix.
This can be easily done in Maxima with the aid of the \texttt{fourier\_elim} package, which
deals with inequalities using the Fourier elimination method\footnote{It is very easy to
do the analysis `by hand' in this case, but we want to illustrate the use of Maxima in general.}.
\begin{verbatim}
(%i20)	H:matrix([-%sigma,%sigma,0],[r+p,-1,0],[0,0,-b])$
(%i21)	eigenvalues(H);
(%o21)	
\end{verbatim}
\begin{align*}
& \left[\left[-\frac{\sqrt{4\sigma (r+p)+(\sigma -1)^2}+\sigma+1}{2},\frac{\sqrt{4\sigma (r+p)+(\sigma -1)^2}-\sigma-1}{2},-b\right]\right., \\
& [1,1,1]\Big]
\end{align*}
\begin{verbatim}
(%i22)	load(fourier_elim)$
(%i23)	fourier_elim(
[4*%sigma*r+4*%sigma*p+%sigma^2-2*%sigma+1>(%sigma +1)^2],[x]
);
(%o23)	[r+p-1>0,%sigma>0] or [-r-p+1>0,-%sigma>0]
\end{verbatim}
Thus, as $\sigma >0$, the eigenvalues have negative real part when $p<1-r$. 
Recalling the original value we are using, $r=28$, we have that $p<-27$ guarantees synchronization. Our matrix could be, then (taking $p=-28$),
$$
\mathbf{H}=\begin{pmatrix}
-10 & \phantom{-}10 & \phantom{-}0 \\
\phantom{-}0 & -1 & \phantom{-}0 \\
\phantom{-}0 & \phantom{-}0 & -8/3
\end{pmatrix}\,,
$$
This, along with \eqref{eq10}, gives
$$
\mathbf{k}=\begin{pmatrix}
0 & 0 & \phantom{-}0 \\
28-z(t) & 0 & -x(t) \\
y(t) & x(t) & \phantom{-}0
\end{pmatrix}
$$
for our coupling matrix. According to \eqref{eq9}, the whole system is (in a simplified notation)
$$
\begin{cases}
\dot{x}=10 (y-x) \\
\dot{y}=28 x -y -x z\\
\dot{z}= xy-8 z/3 \\
\dot{x}_s=10(y_s- x_s) \\
\dot{y}_s=28 x_s -y_s -x_s z_s +(28+z)(x-x_s) -x(z-z_s) \\
\dot{z}_s= x_sy_s-8 z_s/3 +y(x-x_s) +x(y-y_s)\,.
\end{cases}
$$

The following Maxima code explores this synchronization technique, again with two points
given by the initial conditions $(5,5,5)$ and $(-5,-5,-5)$.
\begin{verbatim}
(%i24)	data4:rk([s*(y-x),
            r*x-y-x*z,
            x*y-b*z,
	            s*(ys-xs),
	            r*xs-ys-xs*zs+(r+z)*(x-xs)-x*(z-zs),
	            xs*ys-b*zs+y*(x-xs)+x*(y-ys)],
	            [x,y,z,xs,ys,zs],[5,5,5,-5,-5,-5],[t,0,20,0.01])$
(%i25)	data41:makelist([p[2],p[3],p[4]],p,data4)$
(%i26)	data42:makelist([p[5],p[6],p[7]],p,data4)$
(%i27)	wxanimate_framerate : 10$
(%i28)	wxanimate_draw3d(d,makelist(i,i,1,length(data4),30),
	    dimensions=[600,500],view=[60,60],
    xrange=[-20,20],yrange=[-25,35],zrange=[-5,50],
	    point_type=filled_circle,point_size=0.325,color=orange_red,
	    points(rest(data41,d-length(data4))),
	    point_type=filled_circle,point_size=0.25,color=blue,
	    points(rest(data42,d-length(data4))));
(%o28)
\end{verbatim}
\begin{center}
\begin{animateinline}[controls=true,poster=last]{6}
\multiframe{34}{i=0+1}{
\input{Figures/animation4-\i.tex}
}
\end{animateinline}
\end{center}

A topic we have not touched upon is that of the rate of synchronization, that is, the velocity with
which the two initial trajectories approach each other. It will suffice to notice that both approaches presented here present similar rates.  The main difference between them is that master-slave synchronization 
can be, in principle, applied even to structurally \emph{different} systems.


\section{Synchronization of pendula}\label{pendula}

The phenomenon of synchronization is not exclusive of chaotic systems. Huygens observed in the 
XVII century that two clocks placed on the same wall, or hanging from the same home beam, were 
quickly synchronized, and the interest on this fact has survived up to now, with some recent findings \cite{6b}. It is an important example to which the synchronization
methods just analyzed can be applied.

Given two identical pendula described by \eqref{eq7} with length $\ell =1{.}5$,
a two-way synchronization can be achieved by choosing constant diagonal matrices
$\mathbf{k}_1=\mathbf{k}_2$ with all diagonal elements equal to $0{.}5$, and applying 
\eqref{eq9b}, naming the phase space variables of the first pendulum $(x,y)$ and those of 
the second $(u,v)$. The initial conditions are $(x=0,y=1)$ and $(u=1,v=0)$:

\begin{verbatim}
(%i1)	g:9.8;l:1.5$
(%i2)	pend1:rk([y+0.5*(u-x),-(g/l)*sin(x)+0.5*(v-y),v+0.5*(x-u),
 -(g/l)*sin(u)+0.5*(y-v)],[x,y,u,v],[0,1,1,0],[t,0,10,0.05])$
(%i3)	data11:makelist([p[2],p[3]],p,pend1)$
(%i4)	data12:makelist([p[4],p[5]],p,pend1)$
(%i5)	wxplot2d([[discrete,data11],[discrete,data12]],[legend,false],
[style,[lines,1],[lines,1],[lines,1]],[ylabel," "],[xlabel," "]);
(%o5)
\end{verbatim}
\begin{center}
\input{Figures/animation5-33.tex} 
\end{center}

Let us set up some cosmethic options and do an animation:	

\begin{verbatim}
(%i6)	wxanimate_framerate:5$
(%i7)	with_slider_draw(d,makelist(i,i,1,length(data11),5),
	    point_type = filled_circle,point_size = 0.5,color=blue,
	    points(rest(data11,d-length(data11))),
	    point_type = filled_circle,point_size = 0.5,color=red,
	    points(rest(data12,d-length(data12))))$
(%t7)
\end{verbatim}	 
\begin{center}
\begin{animateinline}[controls=true,poster=last]{6}
\multiframe{34}{i=0+1}{
\input{Figures/animation5-\i.tex}
}
\end{animateinline}
\end{center}

To study the master-slave synchronization of the pendula, notice first that the differential
of the pendulum vector field is
\begin{equation*}
\mathrm{d}_{\mathbf{x}(t)}\mathbf{f} =
\frac{\partial (\dot{x},\dot{y},\dot{z})}{\partial (x,y,z)}=
\begin{pmatrix}
0 & 1 \\
-\frac{g}{l}\cos x & 0
\end{pmatrix}\,.
\end{equation*}

Moreover, we choose the simplest $\mathbf{H}$ Hurwitz, in the form
$$
\begin{pmatrix}
h & 0 \\
0 & h
\end{pmatrix}
$$
with $h<0$. Then, \eqref{choicek} leads to the coupling matrix
$$
\mathbf{k}=\begin{pmatrix}
-h & 1 \\
-\frac{g}{l}\cos x & -h
\end{pmatrix}\,.
$$

The following Maxima session implements this technique.

\begin{verbatim}
(%i8)	h:-0.5$
(%i9)	pend2:rk([y,-(g/l)*sin(x),
 v-h*(x-u)+(y-v),-(g/l)*sin(u)-(g/l)*(x-u)*cos(x)-h*(y-v)], 
 [x,y,u,v],[0,1,0.5,0.5], [t, 0, 10, 0.05])$
(%i10)	data21:makelist([p[2],p[3]],p,pend2)$
(%i11)	data22:makelist([p[4],p[5]],p,pend2)$
(%i12)	wxplot2d([[discrete,data21], [discrete,data22]],[legend,false],
 [style,[lines,1],[lines,1],[lines,1]],[ylabel," "],[xlabel," "]);
(%o12)
\end{verbatim}
\begin{center}
\input{Figures/animation6-33}
\end{center}
The corresponding animation shows clearly the quick synchronization that occurs:
\begin{verbatim}	
(%i13)	with_slider_draw(d,makelist(i,i,1,length(data21),5),
	    point_type = filled_circle,point_size = 0.5,color=blue,
	    points(rest(data21,d-length(data21))),
	    point_type = filled_circle,point_size = 0.5,color=red,
	    points(rest(data22,d-length(data22)))
	)$
(%t13)	
\end{verbatim}
\begin{center}
\begin{animateinline}[controls=true,poster=last]{6}
\multiframe{34}{i=0+1}{
\input{Figures/animation6-\i.tex}
}
\end{animateinline}
\end{center}

As a final remark, let us mention that the synchronization of three or more pendula is
done in a similar way. The reader will experience no difficulties in writing down the 
corresponding commands.


\appendix

\section{Stability and Lyapunov functions}\label{Appendix}

\begin{figure}[h]\centering
\begin{tikzpicture}
 \begin{scope}[domain=-3:3, samples=100]
            \draw [red,thick] plot (\x, {0.09*(\x^3)});
            \draw [gray!75,dashed,thin] plot (\x, {0.09*(\x^3) +0.5});
            \draw [gray!75,dashed,thin] plot (\x, {0.09*(\x^3) -0.5});
            \draw [blue,thick] plot (\x, {0.09*(\x^3) +0.5*sin(3*\x r)});           
 \end{scope}
\draw[stealth-stealth] (0,-0.5)--(0,0.5)node[above]{$2\varepsilon$};       
\end{tikzpicture}\caption{Lyapunov stability}\label{fig4}
\end{figure}
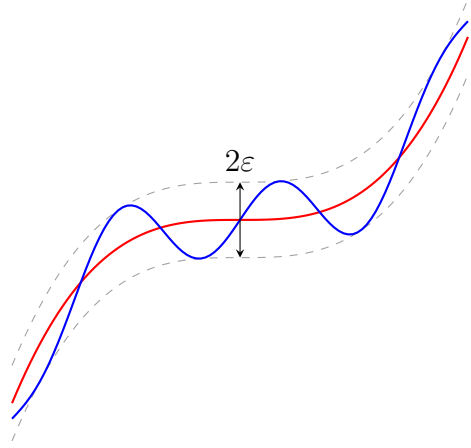

Given a general, non-autonomous dynamical system on $\mathbb{R}^n$
\begin{equation}
\dot{{\bf x}}(t)=\mathbf{f}({\bf x}(t),t)\,,
\label{Lyapunov}
\end{equation}
let us denote its solutions by ${\bf x} =\varphi(t)$, ${\bf x} =\psi(t)$, etc.
A solution ${\bf x}= \varphi(t)$ to \eqref{Lyapunov} is said to be \emph{Lyapunov stable}
if and only if, for all $\varepsilon >0$, there exists a $\delta_{\varepsilon}$
such that for any other solution ${\bf x}= \psi(t)$ satisfying $||\varphi (t_0)-\psi (t_0)||<\delta_{\varepsilon}$ (that is, in a certain instant), then
$$
||\varphi (t)-\psi (t)||<\varepsilon, \;\;\; \forall t>t_0
$$
(that is, for any other later instant). Graphically, we have the situation depicted in Figure
\ref{fig4}.
In other words: A given solution is stable when any other solution, close to it at a
given instant, remains close for any other later instant (that does \emph{not} imply
that both solutions converge, as seen in Figure \ref{fig4} one can circle around the other;
the notion of asymptotic stability given below is related to this intuitive idea of
asymptotic convergence).
On the contrary, if a solution is unstable there is at least another solution that
starts close to it and quickly diverges.

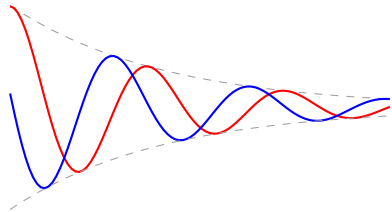
\begin{figure}[h]\centering
\begin{tikzpicture}
 \begin{scope}[scale=2,domain=2:4.5, samples=100]
            \draw [gray!75,dashed,thin] plot (\x, {5*e^(-\x)});
            \draw [red,thick] plot (\x, {5*e^(-\x)*sin(7*\x r)});
            \draw [blue,thick] plot (\x, {5*e^(-\x)*cos(7*\x r)});            
            \draw [gray!75,dashed,thin] plot (\x, {-5*e^(-\x)});            
 \end{scope}   
\end{tikzpicture}\caption{Asymptotic stability}\label{fig5}
\end{figure}

A solution ${\bf x}= \varphi(t)$ to \eqref{Lyapunov} is said to be \emph{asymptotically
stable} if and only if it is stable \emph{and} there exists a $\delta >0$ such that
for any other solution ${\bf x} =\psi (t)$ satisfying $||\varphi (t_0)-\psi (t_0)||<\delta$
then
$$
||\varphi (t)-\psi (t)|| \rightarrow 0, \mbox{ when } t\rightarrow \infty\,.
$$
Thus, a solution $\varphi $ is  asymptotically stable if (aside from being stable)
any other solution $\psi $ staying close to it at a certain instant, gets closer and
closer as $t \rightarrow \infty$. Figure \ref{fig5} offers a graphical description.

Let us remark that, according to the definition, asymptotic stability implies stability, but the
reciprocal is not true. A trivial counter-example is provided by the one-dimensional system
$\dot{x}=0$, whose solutions are all of the form $x(t)=k$, with $k \in \mathbb{R}$ a constant.
Each of these solutions is stable (it suffices to consider
$\delta = \varepsilon$ in the definition), but none of them is asymptotically stable (if a
solution $\varphi (t)$ corresponds to $k_1$, and another solution $\psi (t)$ corresponds to $k_2$,
they always are separated by a distance $||\varphi (t)-\psi (t)||=|k_1 - k_2 |$).

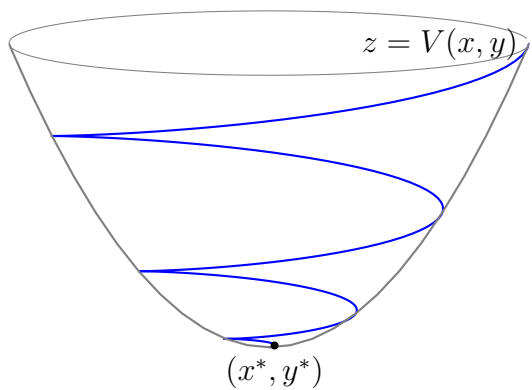
\begin{figure}[h]\centering
\begin{tikzpicture}
\coordinate (O) at (3.5,0.45);
\coordinate (P) at (4.5,4.45);
\begin{axis}[hide axis,view={0}{10},]
\addplot3[y domain=0:0,blue,thick,samples=200,domain=0:3,]
({x*cos(deg(2*pi*x))},{x*sin(deg(2*pi*x))},{x^2});
\addplot3[domain=-3:3,y domain=0:0,gray,thick]({x},{0},{x^2});
\addplot3[domain=-3:3,y domain=0:0,gray,samples=200]({x},{sqrt(9-x^2)},{9});
\addplot3[domain=-3:3,y domain=0:0,gray,samples=200]({x},{-sqrt(9-x^2)},{9});
\fill[color=white] (P)  circle[radius=0.5pt] node[right,color=black]{$z=V(x,y)$};
\fill (O) circle[radius=1.5pt] node[below]{$(x^*,y^*)$};
\end{axis}
\end{tikzpicture}\caption{A Lyapunov function}\label{fig6}
\end{figure}

To explain Lyapunov's technique for proving the stability of a system, consider a plane system described
by the following set of equations:
$$
\begin{cases}
\dot{x}=f(x,y) \\
\dot{y}=g(x,y)
\end{cases}
$$
If $(x^\ast ,y^\ast)$ is an equilibrium point, we know that $f(x^\ast ,y^\ast)=0=g(x^\ast ,y^\ast)$.
Let us study the stability of the constant solution $(x(t),y(t))=(x^\ast ,y^\ast)$, but assuming a
certain set of particular conditions:
\begin{enumerate}
\item\label{lya1} There exists a smooth function (at least $\mathcal{C}^1$) $V(x,y)$ defined on a
neighborhood $D$ of $(x^\ast ,y^\ast)$ which is positive-definite, so its graph lies on the positive
half-space.
\item\label{lya2} The function  $z=V(x,y)$ possesses a unique critical point, a minimum, 
located precisely at $(x^\ast ,y^\ast)$.
\item\label{lya3} The function $V(x,y)$ is decreasing along trajectories of the system; that is: If $z(t)=V(x(t),y(t))$, then
$$
\frac{dz}{dt}\leq 0\,.
$$
\end{enumerate}
A function $V=V(x,y)$ satisfying these conditions is called a \emph{Lyapunov function}. A typical
level surface $z=V(x,y)$ is depicted in Figure \ref{fig6}, along with a trajectory $z(t)=V(x(t),y(t))$.

Lyapunov's theorem states that, under assumptions \ref{lya1}, \ref{lya2}, \ref{lya3} above,
the equilibrium $(x^\ast ,y^\ast)$ is \emph{stable}, and it is \emph{asymptotically stable} if
$$
\frac{dz}{dt}< 0
$$
for all points in the domain $D-\{(x^\ast ,y^\ast)\}$\footnote{Although the present discussion
takes place in $\mathbb{R}^2$ for clarity of exposition, it should be clear that Lyapunov's method
is equally applicable in any $\mathbb{R}^n$ with the appropriate modifications in notation.}.

Condition \ref{lya3} above is perhaps the most important. It implies that, if we run along the
solution $(x(t),y(t))$ starting at $(x_0 ,y_0 )$ for $t=0$, we must `descent' sliding down the
surface $z=V(x,y)$. Because of the profile of this surface, the points $(x(t),y(t))$ in the
descending trajectory must approach $(x^\ast ,y^\ast)$, and this is precisely the condition for
(asymptotic) stability of the equilibrium $(x^\ast ,y^\ast)$.

The main advantage of Lyapunov's method is that, when it is possible to find one, the existence of a
suitable $V$ guarantees asymptotic stability, a property that can be very difficult to prove by 
other means. Moreover, the technique is applicable to systems that are not necessarily linear
nor autonomous.

Although there are no general rules to construct a Lyapunov function for a given system, there
are some general guiding lines that can be useful in particular cases. For instance, a good
strategy consists in trying to find a Lyapunov function quadratic in the system's variables.
This is consistent with the fact that, in many physical problems, $V$ can be interpreted as a
kind of energy of the system (which is dissipated through its motion), and usually the energy
has a quadratic dependence both on positions and velocities.

Let us see an example, using Maxima to do the (somewhat boring) computations.
Consider the unique equilibrium $x^\ast =(0,0)$ of the system
$$
\begin{cases}
\dot{x}=-2xy \\
\dot{y}=x^2 -y^3\,,
\end{cases}
$$
and let us look for a function of the form
$$
E(x,y)=\alpha x^{2p}+\beta y^{2q}
$$
that satisfies Lyapunov's hypothesis.
Notice that we will have to compute the derivative of $E(x(t),y(t))$ with respect to
$t$. In doing so, after applying the chain rule, derivatives such as $\dot{x}=-2xy$
and $\dot{y}=x^2 -y^3\,$ will appear. We can instruct Maxima to consider $x$ and $y$
as functions of $t$, with derivatives given by these functions, with the commands
\texttt{depends} and \texttt{gradef}, respectively.
\begin{verbatim}
(%i1) depends([x,y],[t]);
\end{verbatim}
$$[x\left( t\right) ,y\left( t\right) ]\leqno{\tt (\%o1) }$$
\begin{verbatim}
(%i2) gradef(x(t),-2*x(t)*y(t));
\end{verbatim}
$$x\left( t\right) \leqno{\tt (\%o2) }$$
\begin{verbatim}
(%i3) gradef(y(t),x(t)^2-y(t)^3);
\end{verbatim}
$$y\left( t\right) \leqno{\tt (\%o3) }$$
\begin{verbatim}
(%i4) E(x,y):=%alpha*(x(t))^(2*p)+%beta*(y(t))^(2*q);
\end{verbatim}
$$E\left( x,y\right) :=\alpha\,{x\left( t\right) }^{2\,p}+\beta\,{y\left( t\right) }^{2\,q}\leqno{\tt (\%o4) }$$
Now we are ready to compute the implicit derivative $\frac{dE}{dt}$:
\begin{verbatim}
(%i5) diff(E(x,y),t);
\end{verbatim}
$$2\,\beta\,q\,{y\left( t\right) }^{2\,q-1}\,\left( {x\left( t\right) }^{2}-{y\left( t\right) }^{3}\right) -4\,\alpha\,p\,{x\left( t\right) }^{2\,p}\,y\left( t\right) \leqno{\tt (\%o5) }$$
\begin{verbatim}
(%i6) expand(%);
\end{verbatim}
$$-2\,\beta\,q\,{y\left( t\right) }^{2\,q+2}+2\,\beta\,q\,{x\left( t\right) }^{2}\,{y\left( t\right) }^{2\,q-1}-4\,\alpha\,p\,{x\left( t\right) }^{2\,p}\,y\left( t\right) \leqno{\tt (\%o6) }$$
We want the property $\frac{dE}{dt} < 0$ to be satisfied for points different from $(0,0)$. As
$-2\,\beta\,q\,{y\left( t\right) }^{2\,q+2}$ is clearly negative-definite when $\beta\,q \geq 0$,
we must try to cancel the remaining terms
\begin{verbatim}
(%i7) a(p,q,%alpha,%beta):=
+2*%beta*q*x(t)^2*y(t)^(2*q-1)-4*%alpha*p*x(t)^(2*p)*y(t);
\end{verbatim}
$$a\left( p,q,\alpha,\beta\right) :=2\,\beta\,q\,{x\left( t\right) }^{2}\,{y\left( t\right) }^{2\,q-1}-4\,\alpha\,p\,{x\left( t\right) }^{2\,p}\,y\left( t\right) \leqno{\tt (\%o7)}$$
Taking into account the condition $\beta\,q \geq 0$, by inspection we see that the choice
$p=1=q$, $\alpha=1$, $\beta=2$ is a good one (others are possible, too):
\begin{verbatim}
(%i8) a(1,1,1,2);
\end{verbatim}
$$0\leqno{\tt (\%o8) }$$
Substituting in $E(x,y)$, we get the desired Lyapunov function:
\begin{verbatim}
(%i9) V(x,y)=subst([p=1,q=1,%alpha=1,%beta=2],E(x,y));
\end{verbatim}
$$V\left( x,y\right) =2\,{y\left( t\right) }^{2}+{x\left( t\right) }^{2}\leqno{\tt (\%o9) }$$
Finally, we check that its values along the trajectories of the system are always negative: 
\begin{verbatim}
(%i10) 'diff(V(x,y),t)=
expand(subst([p=1,q=1,%alpha=1,%beta=2],diff(E(x,y),t)));
\end{verbatim}
$$\frac{d}{d\,t}\,V\left( x,y\right) =-4\,{y\left( t\right) }^{4}\leqno{\tt (\%o10) }$$


\end{document}

%% file: Figures/animation1-39.tex
\begin{tikzpicture}
\begin{axis}[axis lines=left,axis line style={->},smooth,scale=1.2,
xmin=-25,xmax=35,xlabel={$x$},
ymin=-25,ymax=35,ylabel={$y$},
zmin=-10,zmax=55,zlabel={$z$}]
\addplot3[thin,Cerulean] table[x="xcoord",y="ycoord",z="zcoord",col sep=comma]{Figures/animation1-39.csv};
\end{axis}
\end{tikzpicture}

%% file: Figures/animation5-33.tex
\begin{tikzpicture}
\begin{axis}[axis lines=left,axis line style={->},smooth,scale=1.2,
xmin=-1,xmax=1,xlabel={$x$},
ymin=-2,ymax=2,ylabel={$y$}]
\addplot[thin,Cerulean] table[x="xcoord",y="ycoord",col sep=comma]{Figures/animation5a-33.csv};
\addplot[opacity=0.66,thin,RedOrange] table[x="xcoord",y="ycoord",col sep=comma]{Figures/animation5b-33.csv};
\end{axis}
\end{tikzpicture}

%% file: Figures/animation6-33.tex
\begin{tikzpicture}
\begin{axis}[axis lines=left,axis line style={->},smooth,scale=1.2,
xmin=-1,xmax=1,xlabel={$x$},
ymin=-2,ymax=2,ylabel={$y$}]
\addplot[thin,Cerulean] table[x="xcoord",y="ycoord",col sep=comma]{Figures/animation6a-33.csv};
\addplot[opacity=0.66,thin,RedOrange] table[x="xcoord",y="ycoord",col sep=comma]{Figures/animation6b-33.csv};
\end{axis}
\end{tikzpicture}

%% file: Synchronization-with-Maxima-arXiv.bbl
\begin{thebibliography}{9}

\bibitem{1} R. L. Devaney: \emph{An Introduction to Chaotic Dynamical Systems} (2nd. Ed.)
Westview Press, 2003.
\bibitem{2} I. Grosu: \emph{Robust Synchronization}. Physical Review \textbf{E56} 3 (1997) 3709--3712.
\bibitem{3} J. K. Hale: \emph{Ordinary Differential Equations}. Dover, 2009.
\bibitem{4} M. W. Hirsch, S. Smale, R. L. Devaney: \emph{Differential Equations, Dynamical Systems, and an Introduction to Chaos} (3rd. Ed.) Academic Press, 2013.
\bibitem{5} E. N. Lorenz: \emph{Deterministic nonperiodic flow}. Journal of the Atmospheric Sciences \textbf{20} 2 (1963) 130--141.
\bibitem{5b} Maxima.sourceforge.net.  Maxima,  a  Computer  Algebra  System.  Version  5.41.0
(2018). \url{http://maxima.sourceforge.net/}.
\bibitem{6} A. Morante, J. A. Vallejo: \emph{Chaotic dynamics with Maxima}. The Electronic Journal
of Mathematics and Technology \textbf{7} 4 (2013) 1--25.
\bibitem{6b} H. M. Oliveira, L. V. Melo: \emph{Huygens synchronization of two clocks}.
Scientific Reports \textbf{5} (2015) 11548.
\bibitem{7} E. Ott: \emph{Chaos in Dynamical Systems} (2nd. Ed.) Cambridge UP, 2002.
\bibitem{8} L. M. Pecora, T. L. Carroll: \emph{Synchronization in Chaotic Systems}.
Physical Review Letters \textbf{64} 8 (1990) 821--824.
\bibitem{9} S. Strogatz: \emph{Nonlinear Dynamics and Chaos} (2nd. Ed.) Westview Press, 2015.
\bibitem{10} C. W. Wu, L. O. Chua: \emph{A simple way to synchronize chaotic systems with applications to secure communication systems}. International Journal of Bifurcation and Chaos 
\textbf{3} 06 (1993) 1619--1627, 1993.
\end{thebibliography}
